\documentclass[twocolumn,showpacs,preprintnumbers,amsmath,amssymb,superscriptaddress]{revtex4}
\usepackage[bookmarks=false]{hyperref}
\usepackage{amssymb}
\usepackage{amsmath}
\usepackage{graphicx}% Include figure files
\usepackage{dcolumn}% Align table columns on decimal point
\usepackage{bm}% bold math
\usepackage{epstopdf}
\usepackage{times}
\usepackage{color}
\begin{document}
\preprint{}
\title
{Magnetic Ordering and Crystal Field Effects in Quasi Caged Structure Compound PrFe$_2$Al$_8$}
\author{Harikrishnan S. Nair}
\email{h.nair.kris@gmail.com, hsnair@uj.ac.za}
\affiliation{Highly Correlated Matter Research Group, Physics Department, P. O. Box 524, University of Johannesburg, Auckland Park 2006, South Africa}
\author{Sarit K. Ghosh}
\affiliation{Highly Correlated Matter Research Group, Physics Department, P. O. Box 524, University of Johannesburg, Auckland Park 2006, South Africa}
\author{Ramesh Kumar K.}
\affiliation{Highly Correlated Matter Research Group, Physics Department, P. O. Box 524, University of Johannesburg, Auckland Park 2006, South Africa}
\author{Andr\'{e} M. Strydom}
\affiliation{Highly Correlated Matter Research Group, Physics Department, P. O. Box 524, University of Johannesburg, Auckland Park 2006, South Africa}
%
%\author{Ramesh Kumar K.}
%\affiliation{Highly Correlated Matter Research Group, Physics Department, P. O. Box 524, University of Johannesburg, Auckland Park 2006, South Africa}
%
\date{\today}
\begin{abstract}
The compound PrFe$_2$Al$_8$ possesses a three-dimensional network structure resulting from the packing of Al polyhedra centered at the transition metal element Fe and the rare earth Pr. Along the $c$-axis, Fe and Pr form {\em chains} which are separated from each other by the Al-network. In this paper, the magnetism and crystalline electric field effects in PrFe$_2$Al$_8$ are investigated through the analysis of magnetization and specific heat data. A magnetic phase transition in the Pr lattice is identified at $T^{Pr}_{N}\approx$ 4~K in dc magnetization and ac susceptibility data. At 2~K, the magnetization isotherm presents a ferromagnetic saturation, however, failing to reach full spin-only ferromagnetic moment of Pr$^{3+}$. Metamagnetic step-like low-field features are present in the magnetization curve at 2~K which is shown to shift upon field-cooling the material. Arrott plots centered around $T^{Pr}_{N}$ display "S"-like features suggestive of an inhomogeneous magnetic state. The magnetic entropy, $S_m$, estimated from specific heat outputs a value of $R$ ln(2) at $T_{N2}$ suggesting a doublet state for Pr$^{3+}$. The magnetic specific heat is modeled by using a 9-level Schottky equation pertinent to the Pr$^{3+}$ ion with $J$ = 4. Given the crystalline electric field situation of Pr$^{3+}$, the inference of a doublet state from specific heat and consequent long-range magnetic order is an unexpected result.
\end{abstract}
\maketitle
\section{\label{INTRO}Introduction}
\indent
Intermetallic alloys of the 1:2:8 composition, $RT_2X_8$, 
possess quasi cage structures consisting of polyhedra that cover the
rare earth and the transition metal. These structural peculiarities of 
cage-like structures are interesting from the perspective of 
thermoelectric properties\cite{koterlin1989bs}
for realizing high thermoelectric power. For example, a positive 
maximum of 20~$\mu$V/K was observed
for CeFe$_2$Al$_8$ at 150~K. However, limited work on such 
compounds and the lack of complete understanding of the observed magnetic properties
are compelling reasons to investigate new compounds in $RT_2X_8$ or
to re-investigate some of the reported ones. The $RT_2X_8$ compounds crystallize in 
orthorhombic space group $Pbam$ ($\#$55) \cite{poettgen_zaac_2001new}. 
In this space group, there are four formula units in the unit cell with one 
Wyckoff position ($4g$) for the $R$ atom, two ($4g$) for $T$ and nine 
($2a$, $2d$, $4g$ and $4h$) for the $X$ atom
\cite{gladishevskii_sovphys_1983,poettgen_zaac_2001new}. 
The $R$ or $T$ atom in this structure is surrounded by nine to 
twelve $X$ atoms. The $R$ and the $T$ atoms form a {\em chain} parallel 
to the $c$-axis which are separated by the $X$ atoms. It is important 
to mention here a major question regarding 
the structural chemistry of 1:2:8 compounds regarding why they do not 
form 2:1:8 or 1:1:5 stoichiometries which are more stable, especially known for 
the Ce-variants\cite{thompson_physica_2003}. It has been found 
from previous studies that the reason lies in the ionic size of the $R$ and the $X$. 
In the case of 1:1:5 and 2:1:8-stoichiometries, the $R$-ion is in the trivalent $f^1$ (Ce)
state whereas in 1:2:8 systems, the rare earth is in divalent or mixed-valent state
\cite{fritsch_prb_2004antiferromagnetic}. 
In Ce-based 1:2:8 systems,
\cite{kolenda_jac_2001low,koterlin1989bs,ghosh_actapolonica_2012strongly} 
a transition from magnetic trivalent state of Ce to non-magnetic 
Kondo-like state of Ce is observed depending on the type of $T$ and $X$ atoms. 
\\
\begin{figure}[!b]
\centering
\includegraphics[scale=0.055]{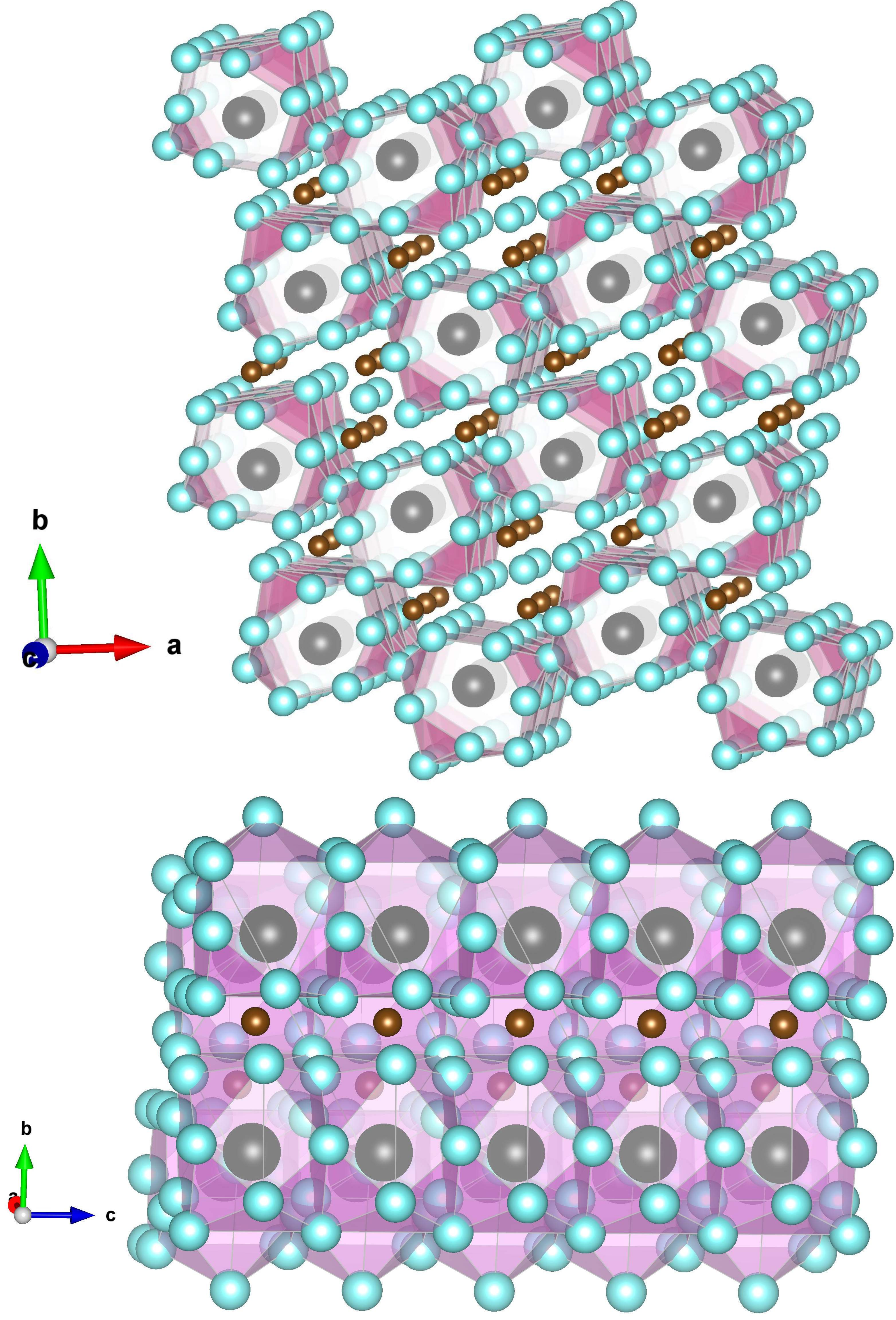}
\caption{\label{fig_str} {\em top:} A projection of the crystal structure of PrFe$_2$Al$_8$ along the $c$-axis showing the network of Al atoms (cyan) and the vacant space where Pr (black) and Fe (brown) form {\em chains} parallel to the $c$-axis. {\em bottom:} The view parallel to $c$-axis showing the chains of Pr. The figure was created using the software VESTA.\cite{vesta}}
\end{figure}
\indent
Strongly correlated electron behaviour was one of the focus 
themes to investigate the $RT_2X_8$ compounds 
and such signatures have been experimentally observed in Ce$T_2$Al$_8$ 
[$T$ = Fe, Co]\cite{ghosh_actapolonica_2012strongly}.
It was argued that the ground state of these compounds was governed by 
RKKY interaction\cite{van1962note} but possessed contrasting 
outcomes -- in one case a metallic Fermi liquid state was evidenced whereas 
in the other, a heavy Fermi-liquid with 
proximity to long-range magnetic order\cite{ghosh_actapolonica_2012strongly}.
Cerium based $RT_2X_8$ with $X$ = Al, Ga have been reported previously also
\cite{sichevich1985intermetallic,koterlin1989bs} on 
which M\"{o}{\ss}bauer spectroscopy\cite{koterlin1989bs} and 
neutron diffraction studies\cite{kolenda_jac_2001low} were performed in order to 
understand the magnetic ground state. Interestingly, no signature of long-range 
magnetic order was obtained in the case of
CeFe$_2$Al$_8$ from neutron powder diffraction data though the magnetic 
susceptibility displayed a broad transition around 5~K
\cite{kolenda_jac_2001low}. Though Ce-based $RT_2$Al$_8$ were investigated 
in detail in the past, only 
a few reports on Pr-based systems exist. For example, PrCo$_2$Al$_8$ 
displays a phase transition at 5~K and metamagnetic-like steps in magnetic 
hysteresis\cite{tougait_jssc_2005prco}. 
From the analysis of high temperature magnetic susceptibility and estimates 
of effective paramagnetic moment, any contribution from Co magnetism was ruled out. 
The absence of magnetism of the $T$ sub-lattice 
was a feature observed in the case of CeFe$_2$Al$_8$ also
\cite{ghosh_actapolonica_2012strongly}. 
In the case of CeCo$_{2-x}T_x$Al$_8$ single crystals where doping 
at the Co-site was performed with $T$ = 
Mn, Fe and Ni, only a slight enhancement of magnetic moment 
compared to the Ce$^{3+}$-only 
moment was observed\cite{treadwell_ic_2014investigation}. It is, hence, 
interesting to investigate other compounds in the 
$RT_2$Al$_8$ class to probe the extent of magnetic ordering in the 
rare earth and transition metal lattices.
\\
\begin{table}[!t]
\centering
\caption{\label{tab1} The atomic positions of PrFe$_2$Al$_8$ in $Pbam$ space group. The refined lattice parameters are $a$ (\AA) = 12.5098(5), $b$ (\AA) = 14.4410(1), and $c$ (\AA) = 4.0401(7). The $U_{iso}$ values were fixed to the values of PrCo$_2$Al$_8$\cite{tougait_jssc_2005prco}.}
\setlength{\tabcolsep}{9pt}
\begin{tabular}{cccccc}\hline\hline
Atom & site & $x$  & $y$  &    $z$   & $U_{iso}$ \\ \hline\hline  
Pr   & $4g$  & 0.3394   & 0.3173 & 0.0 & 0.0029 \\
Fe(1)   & $4g$  & 0.1511  & 0.0951 & 0.0 & 0.0018 \\
Fe(2)   & $4g$  & 0.0291 & 0.3969 & 0.0 & 0.0021 \\
Al(1)   & $2d$  & 0.0000  & 0.5000 & 0.5 & 0.0033 \\
Al(2)   & $2a$  & 0.0000  & 0.0000 & 0.0 & 0.0053 \\
Al(3)   & $4h$  & 0.0264  & 0.1376 & 0.5 & 0.0053 \\
Al(4)   & $4h$  & 0.1627  & 0.3703 & 0.5 & 0.0031 \\
Al(5)   & $4h$  & 0.2313  & 0.1663 & 0.5 & 0.0035 \\
Al(6)   & $4h$  & 0.3389  & 0.4818 & 0.5 & 0.0034 \\
Al(7)   & $4h$  & 0.4598  & 0.1834 & 0.5 & 0.0028 \\
Al(8)   & $4g$  & 0.3347  & 0.0371  & 0.0 & 0.0041 \\
Al(9)   & $4g$  & 0.0953  & 0.2545  & 0.0 & 0.0043 \\ \hline\hline
\end{tabular}
\end{table}
% %
%
\begin{figure}[!b]
\centering
\includegraphics[scale=0.48]{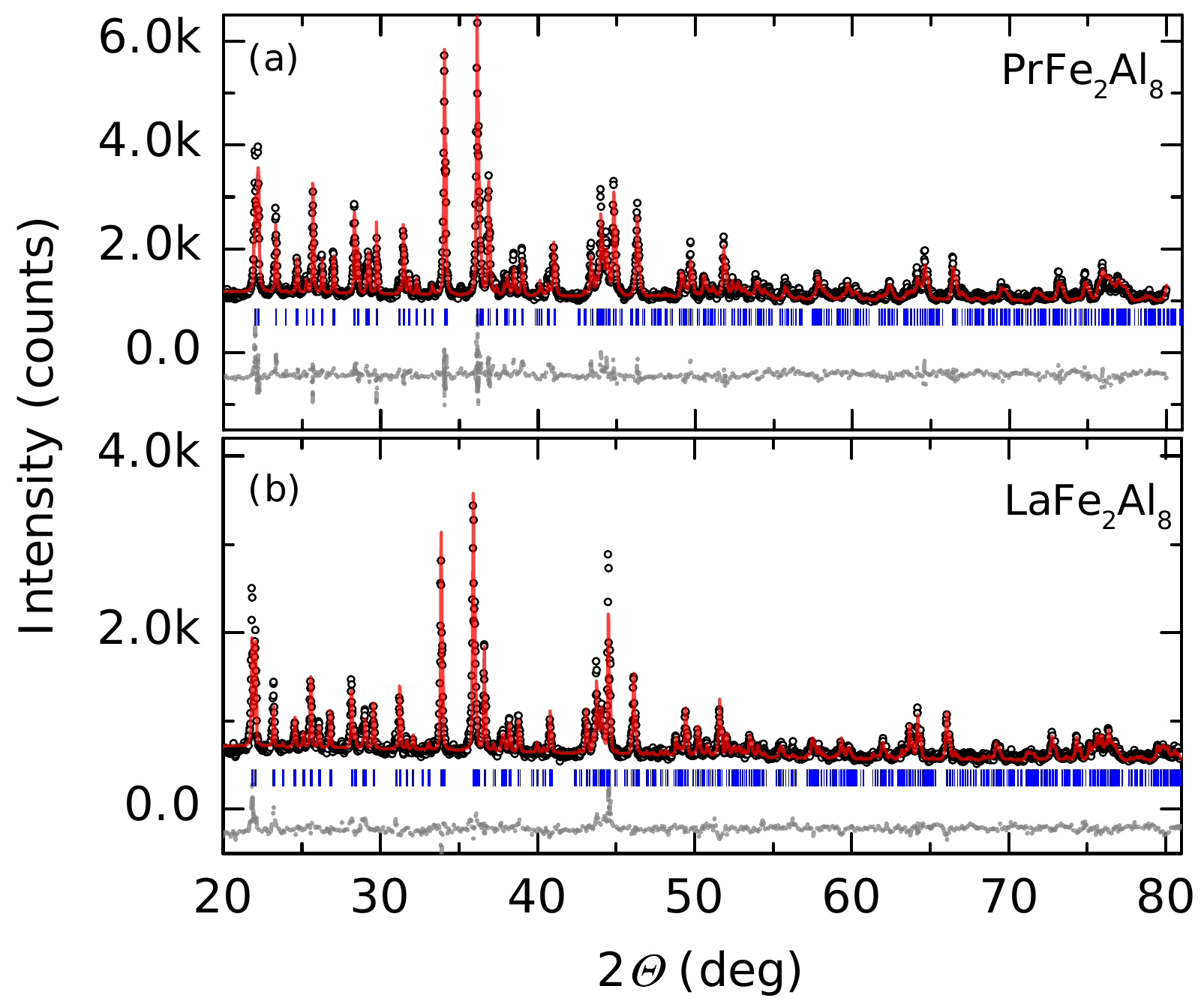}
\caption{\label{fig_xrd} The experimentally observed x ray powder diffraction pattern of PrFe$_2$Al$_8$ shown in black circles in (a). The red solid line is the Rietveld fit assuming $Pbam$ space group model. The vertical bars are the allowed Bragg peaks and the gray dotted line is the difference curve. For a comparison, the x ray diffraction pattern of the isostructural compound LaFe$_2$Al$_8$ is presented in (b) along with the results of Rietveld refinement.}
\end{figure}
\indent
In the present paper we investigate the magnetism of the 
Pr-based $RT_2X_8$ compound PrFe$_2$Al$_8$. 
Fig.~\ref{fig_str} (top) shows the projection of the crystal structure on the 
$ab$-plane displaying the network of Al and the chain-like
formation of Pr and Fe along the $c$-axis (bottom). These structural peculiarities 
certainly have a
bearing on the magnetism of this class of compounds. The magnetism of 
Pr-based compounds
can be strongly influenced by the crystal field levels of Pr$^{3+}$. 
Depending on whether the crystal field levels are comparable in
energy scales with other interactions (especially magnetic), non-magnetic
singlet or doublet crystal field ground states can result. To cite an 
example, the point charge model calculations 
for the Pr site in PrSi predicted the splitting of $J$ = 4 Pr$^{3+}$ 
ground state in to 9 singlets thereby
rendering spontaneous magnetic order impossible. However, 
an anomalous
\begin{figure}[!b]
\centering
\includegraphics[scale=0.28]{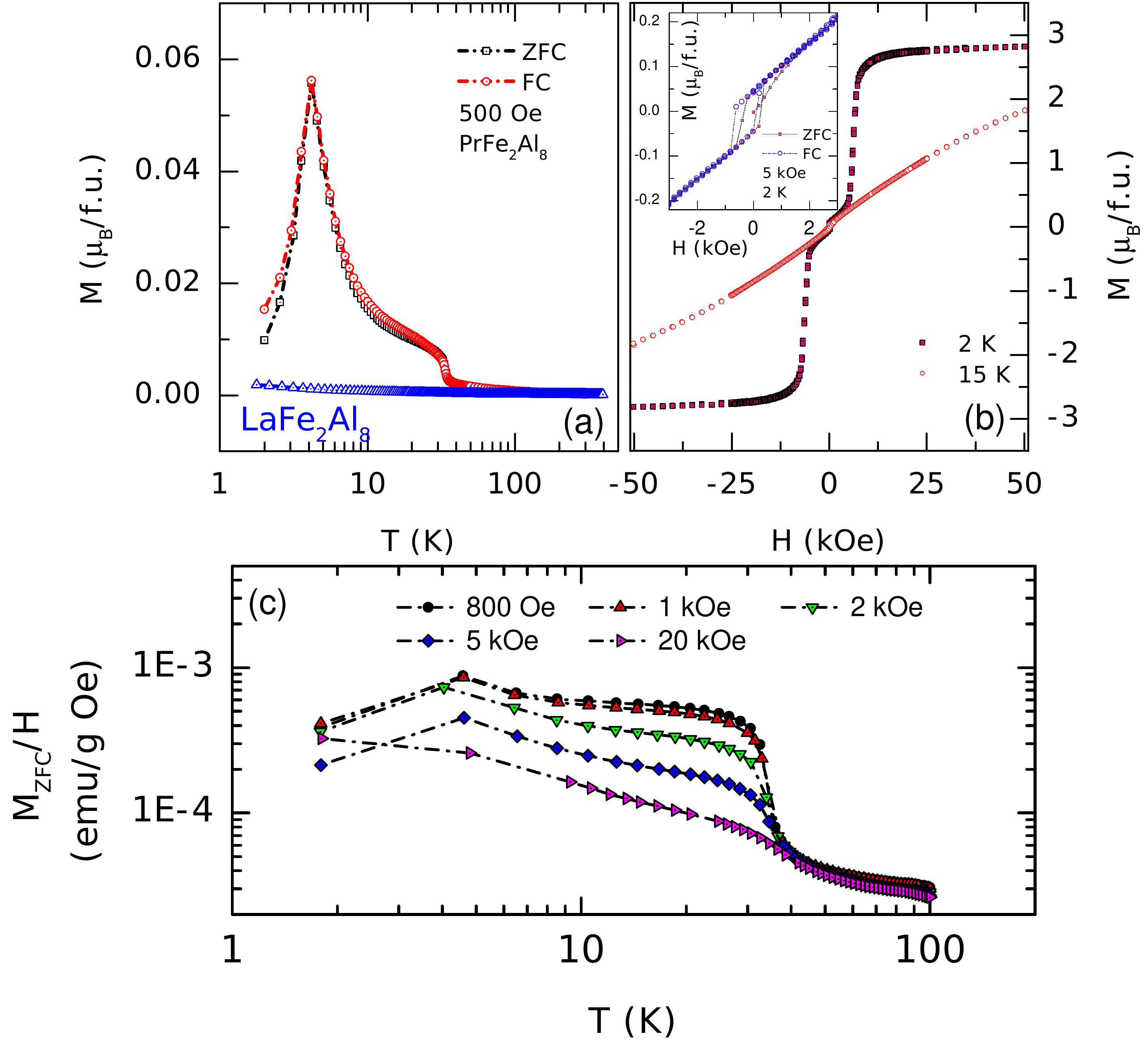}
\caption{\label{fig_mt} The magnetization curves of PrFe$_2$Al$_8$ in ZFC and FC modes are shown in (a). Two transitions at $T_{anom} \approx$ 34~K and at $T^{Pr}_{N} \approx$ 4~K are observed. The magnetization curve of LaFe$_2$Al$_8$ is also presented for a comparison. (b) Shows the magnetic hysteresis at 2~K and at 15~K. Step-like features are observed in the low-field region at 2~K shown magnified in the inset. The inset also present field-cooled data exhibiting an exchange bias-like shift. (c) Displays $M_\mathrm{ZFC}/H$ at different applied fields 800~Oe to 20~kOe. The transition at $T_{N2}$ is suppressed at 10~kOe while that at $T_{anom}$ appears very robust.}
\end{figure}
ferromagnetic ground state was experimentally unraveled in 
PrSi\cite{snyman2012anomalous}.
Similarly, the point symmetry of the Pr-site in PrFe$_2$Al$_8$ is $m$ ($C_s$)
which predicts 9 distinct singlets from a 9-fold degenerate state 
and presupposes a non-magnetic ground state. 
However, through magnetization measurements we observe 
magnetic order in the Pr-lattice which is 
also supported by specific heat data and analysis. 
A crystal field quasi-doublet for Pr$^{3+}$ at low temperature 
could be deduced in this compound.
\section{\label{EXP}Experimental details}
\indent
Polycrystalline samples of PrFe$_2$Al$_8$ were prepared using 
arc melting method starting with high 
purity elements Pr (4$N$), Fe (3$N$) and Al (5$N$). The melting of 
stoichiometric amounts of materials were carried out
in the water-cooled Cu-hearth of a Edmund Buehler furnace in static 
Argon atmosphere. A Zr-getter trap was used for purifying the Ar gas. 
The melted samples were annealed at 900~$^\circ$C for 14 days. Powder x ray diffraction 
patterns were recorded using a Philips X'pert diffractometer which 
employed Cu-K$\alpha$ radiation. Magnetic properties were measured 
in the temperature range of 2 -- 300~K using a Magnetic Property 
Measurement System
(Quantum Design Inc. San Diego), in which magnetization in the zero 
field cooled (ZFC) and field cooled (FC) cycles, magnetic hysteresis 
and ac susceptibility were measured. Specific heat 
was measured in the temperature range 2 -- 300~K using a commercial 
Physical Property Measurement System (also Quantum Design).
\begin{table}[!t]
\centering
\caption{\label{tab2} The interatomic distances in PrFe$_2$Al$_8$ compared with those in LaFe$_2$Al$_8$. The structures of both the compounds were refined in $Pbam$ space group. The distances given in the table are in the units of Angstrom ($\AA$). In all cases, a range of distances are indicated which is due to the quasi-caged structure of these compounds. The Pr-Pr distance along the $c$-axis is given in the table. Along $a$ and $b$, the Pr-Pr distances vary in the range of 6.5 to 7.5~$\AA$.}
\setlength{\tabcolsep}{9pt}
\begin{tabular}{lll}\hline\hline
PrFe$_2$Al$_8$   &  LaFe$_2$Al$_8$ \\ \hline\hline
Pr-Pr: 4.04    & La-La: 4.06 \\
Fe(1)-Fe(1): 4.04 -- 6.18 & Fe(1)-Fe(1): 4.06 -- 6.21 \\
Fe(2)-Fe(2): 3.06 -- 4.04 & Fe(1)-Fe(1): 3.07 -- 4.06 \\
Fe(1)-Fe(2): 4.61 -- 6.36 & Fe(1)-Fe(1): 4.63 -- 6.40 \\
Fe-Al: 2.21 -- 2.64 & Fe-Al: 2.45 -- 2.66 \\
Pr-Al: 3.09 -- 3.36  & La-Al: 3.13 -- 3.20 \\ \hline\hline
\end{tabular}
\end{table}
\begin{figure}[!b]
\centering
\includegraphics[scale=0.31]{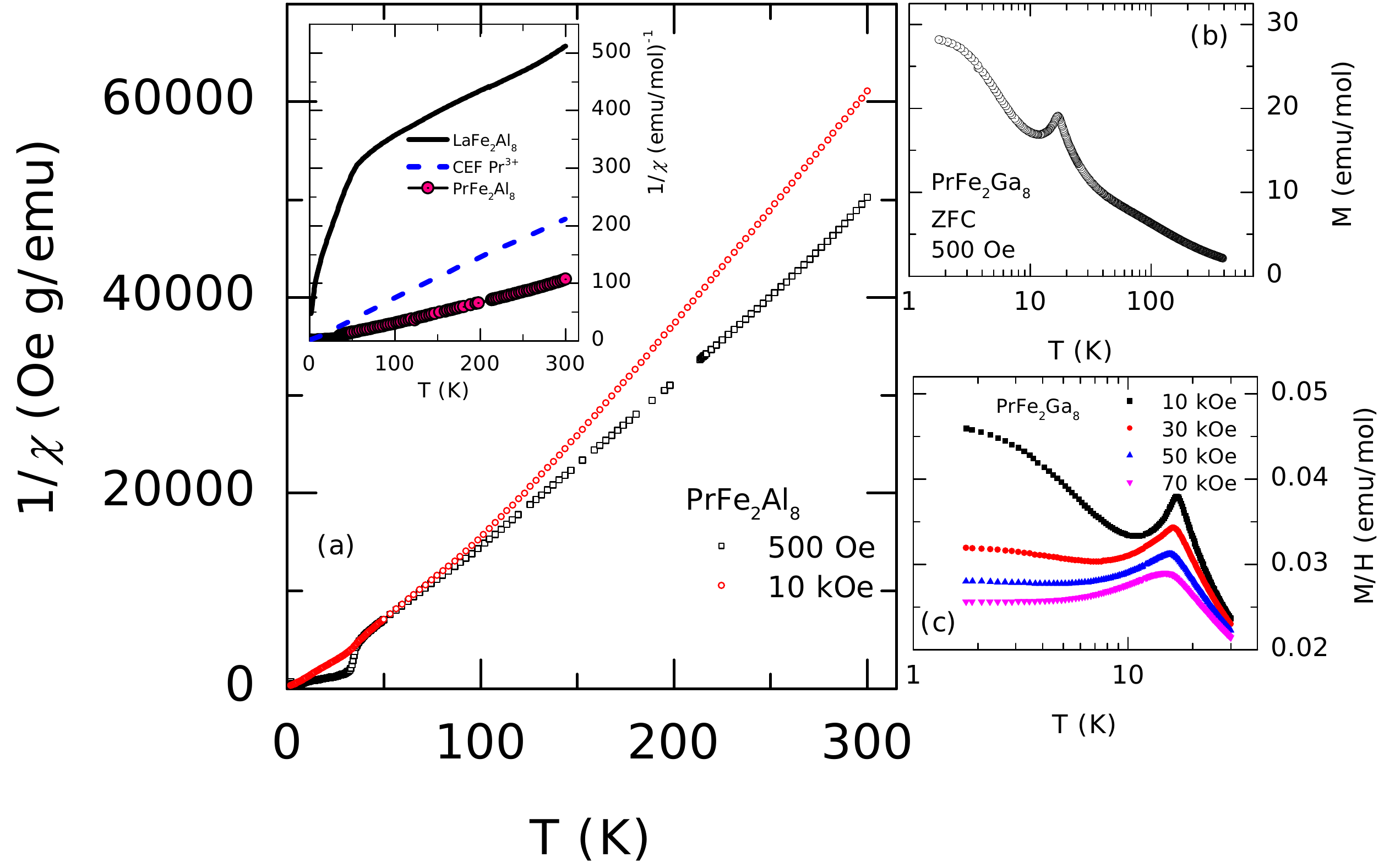}
\caption{\label{fig_cw} (a) The inverse magnetic susceptibility, 1/$\chi(T)$, of PrFe$_2$Al$_8$ plotted for two different applied fields 500~Oe and 10~kOe. The 1/$\chi(T)$ plots are non-linear and the two curves do not scale suggesting magnetic correlations in the Fe lattice. In the inset, the 1/$\chi(T)$ of PrFe$_2$Al$_8$ is plotted along with that of LaFe$_2$Al$_8$. Also shown is the theoretically estimated CEF contribution to total magnetic susceptibility of PrFe$_2$Al$_8$. (b) The magnetization curve of isostructural and isoelectronic compound PrFe$_2$Ga$_8$ shows only one magnetic transition at $\approx$ 15~K. The effect of applied magnetic field is to shift the peak towards lower temperature however the peak is not diminished in strength as can be seen from (c).}
\end{figure}
\section{\label{RESULTS}Results and Discussion}
\subsection{\label{Crys} Crystal structure}
\indent 
The experimentally obtained powder x ray diffraction pattern of 
PrFe$_2$Al$_8$, shown in Fig.~\ref{fig_xrd} (a), 
was refined in the orthorhombic space group $Pbam$. The structural 
parameters of PrCo$_2$Al$_8$\cite{tougait_jssc_2005prco}
were used as the starting model for the Rietveld refinement
\cite{rietveld} using FullProf software\cite{fullprof}.
The results of the refinement procedure are presented in (a) as 
solid lines along with the allowed Bragg peaks (vertical ticks) 
and the difference pattern (dotted line). The refined atomic
and lattice parameters are collected in Table~\ref{tab1}. In 
Fig.~\ref{fig_xrd} (b), the diffraction pattern
of LaFe$_2$Al$_8$ is presented along with the results of Rietveld fit. 
A comparison of the different bond distances of both the 
compounds are collected in Table~\ref{tab2}.
The typical Pr--Al distances in PrFe$_2$Al$_8$ are found to be 
in the range of $3.09~$\AA~to $3.36~$\AA, which is comparable
to the distances observed for the Eu-pentagonal prism in 
EuRh$_2$Ga$_8$\cite{fritsch_prb_2004antiferromagnetic}. 
The Pr--Pr distances are found to be in the range 
$4.04~$\AA~to $7.55~$\AA. 
The Fe atoms have two crystallographic positions Fe(1) and Fe(2). 
The Fe(1)--Fe(1) distances are in the 
range $4.04~$\AA~to $6.18~$\AA~and the Fe(2)--Fe(2) distances 
are $3.06~$\AA~to $4.04~$\AA. 
The Fe--Al distances are $2.21~$\AA~to $2.64~$\AA. These distances 
are comparable to those found in LaFe$_2$Al$_8$
except for the larger La--La distances. In the case of CeFe$_2$Al$_8$, 
the Fe(1)--Fe(1) and the Fe(2)--Fe(2) distances were reported to 
be $7.6$ and $2.8~$\AA~respectively\cite{kolenda_jac_2001low}. 
Given that the Fe--Fe distances in PrFe$_2$Al$_8$ and 
LaFe$_2$Al$_8$ are large and that they are separated by the 
Al-network, it is unlikely that the 
Fe-lattice would have direct magnetic exchange to lead to the 
development of long-range magnetic order. A careful comparison
and correlation of the structural details like bond distances and 
Fe--Fe separation in Pr-based 1:2:8 compounds with the
observed magnetic properties is very much called for.
\begin{figure}[!b]
\centering
\includegraphics[scale=0.35]{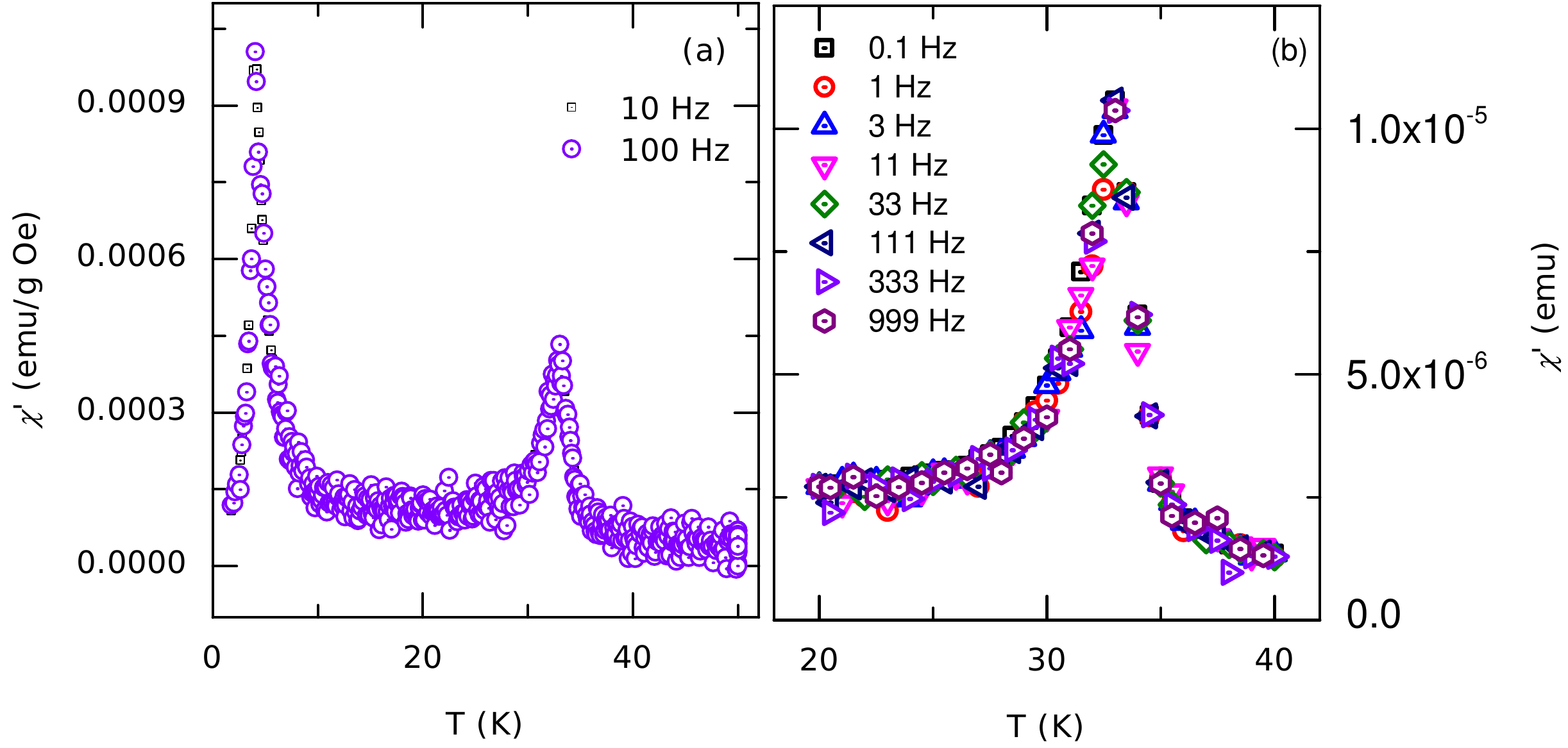}
\caption{\label{fig_chi} (a) The real part of the ac susceptibility, $\chi'(T)$, in the temperature range 2 -- 60~K showing anomalies at $T_{anom}$ and $T^{Pr}_{N}$ with applied frequencies 10~Hz and 100~Hz. (b) The $\chi'(T)$ at $T_{anom}$ is measured for wider range of frequencies, 0.1 -- 1000~Hz, in order to probe the magnetic ordering in detail. Neither transition displays frequency-dependent shift of the peak thereby ruling out any dynamic short-range magnetic order.}
\end{figure}
\subsection{\label{Mag} Magnetic properties}
\indent 
The magnetic response, $M (T)$, of PrFe$_2$Al$_8$ is plotted 
in Fig.~\ref{fig_mt} (a) where the ZFC/FC curves of magnetization at 
applied field of 500~Oe are presented. Two anomalies can be 
clearly distinguished in the magnetic response. As temperature is
decreased from 300~K, the first anomaly is observed as a 
shoulder-like feature at $T_{anom} \approx$ 34~K while the 
second one manifests as a sharp peak at $T^{Pr}_{N}\approx$ 4~K. 
Here we tentatively argue that the latter anomaly pertains 
to the magnetic phase transition
occurring in the Pr sublattice. The anomaly at $T_{anom}$ 
could be due to short-range ordering in the 
Fe sublattice or even due to an impurity phase which was 
not detectable in the x ray diffraction data.
A weak bifurcation between the ZFC and the FC curves is 
observed immediately close to the $T_{anom}$. Also shown 
in (a) is the plot
of magnetization of the isostructural and non-magnetic 
analogue LaFe$_2$Al$_8$,
in which Fe is in principle the only magnetic species, which 
displays no magnetic transition in the temperature range 2--300~K.
On the other hand, PrCo$_2$Al$_8$ displayed a magnetic 
transition
at $T_N \approx$ 5~K where Pr enters a magnetically long-range 
ordered state\cite{tougait_jssc_2005prco}.
No signature of magnetism of the Co-lattice was observed. 
Similar absence of magnetism of 
the transition metal was reported in the case of 
CeFe$_2$Al$_8$ where
intermediate valence of Ce ion leading to valence-fluctuation 
scenario was proposed\cite{kolenda_jac_2001low}.
Through neutron powder diffraction and magnetic studies 
clustering of the Fe ions in CeFe$_2$Al$_8$ was 
suggested\cite{kolenda_jac_2001low}.  The structural similarity 
and comparable interatomic distances
of Fe--Fe of PrFe$_2$Al$_8$ to other 1:2:8 compounds 
(LaFe$_2$Al$_8$, PrCo$_2$Al$_8$ etc) 
might lead to an assumption that the Fe-lattice remains 
non-magnetic in PrFe$_2$Al$_8$.
\\
\begin{figure}[!t]
\centering
\includegraphics[scale=0.30]{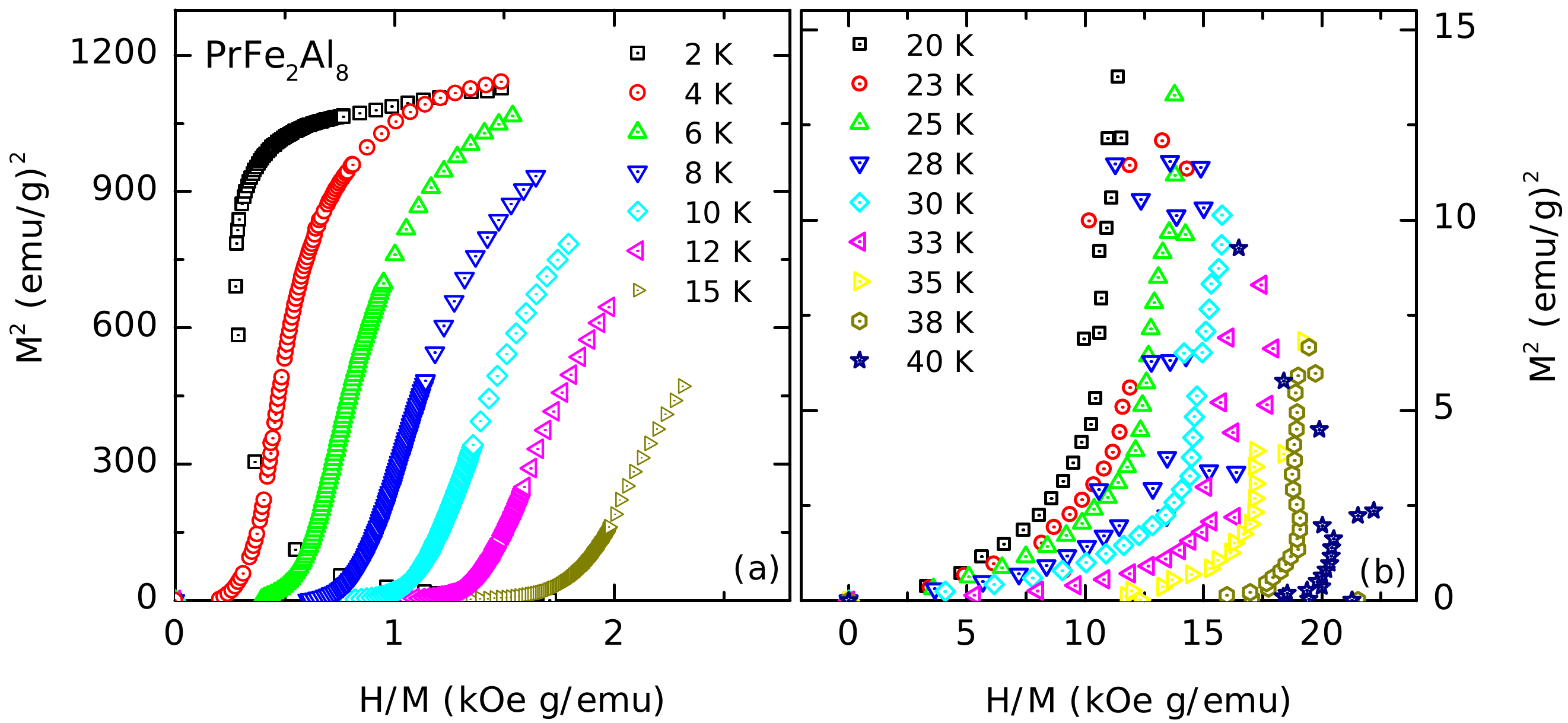}
\caption{\label{fig_arrott} (a) The Arrott plots of PrFe$_2$Al$_8$ derived from the magnetization isotherms between 2~K and 15~K which covers the temperature range around $T_{N2}$. The Arrott plots for the range 20~K to 40~K across $T_{anom}$ is shown in (b). No signature of spontaneous ferromagnetism is observed in either cases however, (b) matches with the case of a disordered ferromagnet.}
\end{figure}
\indent
Figure~\ref{fig_mt} (b) shows the results of field-scans of 
magnetization, $M(H)$, performed at 2~K and at 15~K. 
At 2~K, a ferromagnetic-like 
saturation of magnetization is obtained for PrFe$_2$Al$_8$. 
The maximum moment attained is 
$M^\mathrm{50~kOe}_{max} \approx$ 3~$\mu_\mathrm B$ 
which is diminished compared to the full ferromagnetic moment 
of free ion Pr$^{3+}$ (3.58~$\mu_\mathrm{B}$). This 
indicates that the Pr-magnetic lattice does not attain 
complete ferromagnetic polarization at 2~K. The reduction 
in magnetic moment might result from short-range magnetism
or cluster formation happening in the Fe sub-lattice, due to 
minor impurity effects in the polycrystalline sample or because of 
the Fe moments anti-aligning once the Pr moments start to 
order magnetically.
In the low-field region of the magnetic hysteresis at 2~K, a 
sharp step-like feature is observed in PrFe$_2$Al$_8$
(inset of Fig.~\ref{fig_mt} (b)). This is reminiscent of the 
metamagnetic 
steps observed in PrCo$_2$Al$_8$ at a critical field of 0.9~T 
at 1.7~K\cite{tougait_jssc_2005prco}. The inset also displays the
magnetic hysteresis obtained in field-cooled (at 5000~Oe) 
conditions at 2~K. A 
shift of the hysteresis loop is present as seen in the inset of (b). 
Though a weak feature, this hints at an inherent exchange 
bias-like effect in PrFe$_2$Al$_8$ \cite{nogues_jmmm_1999exchange}. 
Field-cooling the sample in 5000~Oe
did not lead to an enhancement of maximum moment value 
attained at 2~K.
The magnetization isotherm at 15~K does not present a 
magnetic 
hysteresis loop or any other features typical of a 
ferromagnet (see Fig.~\ref{fig_mt} (b)).
In Fig.~\ref{fig_mt} (c), the zero field-cooled magnetization 
curves obtained on PrFe$_2$Al$_8$ with different applied fields 
(800~Oe to 20~kOe) are plotted
together as $M_\mathrm{ZFC}/H$. It can be seen that the magnetic transition at $T^{Pr}_{N}$ 
is suppressed, however, the anomaly at $T_{anom}$ is robust 
against 20~kOe. The peaks of the transitions are seen to remain without any 
significant shift upon the application of magnetic field.
\\
\begin{figure}[!t]
\centering
\includegraphics[scale=0.48]{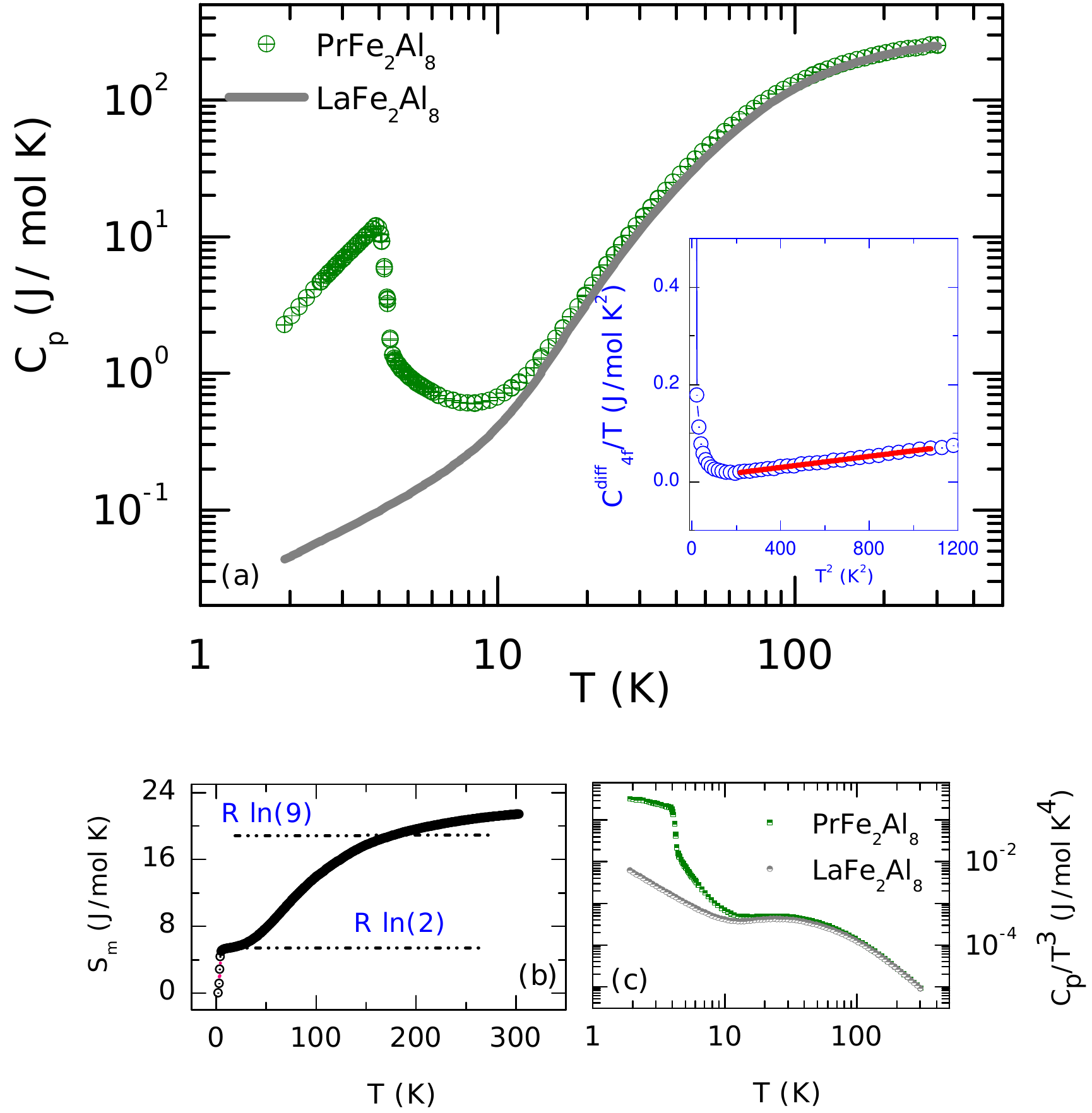}
\caption{\label{fig_cp} (a) The specific heat, $C_p(T)$, of PrFe$_2$Al$_8$ is presented in circles. The phase transition at $T_{N2} \approx$ 4~K is evident. The solid line is the $C_p(T)$ of the non-magnetic analogue, LaFe$_2$Al$_8$. The inset shows the plot of $C_{4f}/T$ versus $T^2$ (circles) and the linear fit (solid line) to obtain $\gamma$. (b) Shows the magnetic entropy, $S_m(T)$ which attains $R$ln(2) at $T_{N2}$. (c) Shows the plot of $C_p/T^3$ versus $T$ to expose the presence of phonon modes.}
\end{figure}
\indent
Fig.~\ref{fig_cw} (a) shows the inverse magnetic susceptibility, 
1/$\chi(T)$, of PrFe$_2$Al$_8$ plotted for two values of applied magnetic fields
{\em viz.,} 500~Oe and 10~kOe. A fit to the data assuming ideal 
Curie-Weiss behaviour was not possible due to the nonlinear
behaviour of 1/$\chi(T)$ especially at elevated temperatures. 
The non-linearity is attributed to crystalline electric field (CEF) effects of Pr.
From (a) it is clear that the 1/$\chi(T)$ curves for different applied
magnetic fields do not scale with each other. This feature is a 
strong indicator to the presence of
short-range magnetic correlations in the Fe-lattice. Curie-Weiss 
fit performed on the 1/$\chi(T)$ data of LaFe$_2$Al$_8$ 
(presented in Fig.~\ref{fig_cw}) resulted in a effective moment 
of $\approx$ 3~$\mu_\mathrm B$. A very high Curie-Weiss 
temperature, $\theta_\mathrm{cw} \approx$ -285~K
was obtained for LaFe$_2$Al$_8$ which suggests prominent antiferromagnetic
correlations obviously stemming from the short-range correlations 
of the Fe subsystem. On the other hand, the non-Fe compound 
PrCo$_2$Al$_8$ is reported to display a linear 1/$\chi(T)$ 
with a perfect Curie-Weiss behaviour and no magnetic contribution 
from Co\cite{tougait_jssc_2005prco}. 
The only magnetic ordering in PrCo$_2$Al$_8$ was detected in the Pr lattice.
In order to disentangle the magnetism of Pr and Fe in PrFe$_2$Al$_8$, 
a rough estimate of
the van Vleck contribution due to Pr$^{3+}$ was theoretically 
calculated
using the software McPhase\cite{mcphase} assuming the Pr$^{3+}$ 
crystalline electric field environment in $Pbam$. 
The theoretically estimated 1/$\chi(T)$ is plotted as a dashed-dotted 
line in the inset of Fig.~\ref{fig_cw}. From this figure
it is clear that in both PrFe$_2$Al$_8$ and LaFe$_2$Al$_8$, 
short-range
correlation of Fe is present till elevated temperatures.
\\
\indent
As a means to understand the multiple magnetic ordering taking place
in PrFe$_2$Al$_8$, the magnetic properties of isostructural, isoelectronic
compound PrFe$_2$Ga$_8$ were also recorded. Those are presented in Fig.~\ref{fig_cw}
(b) and (c). PrFe$_2$Ga$_8$ presents only one magnetic transition at $\approx$
15~K. At very low temperatures close to 2~K, a saturation-like effect is discernible which hints at
possible imminent magnetic ordering phenomena. The effect of applied magnetic fields
on the transition temperatures is probed and the data is presented in (c). It is observed
that the peak at 15~K shifts to low temperatures upon application of magnetic field.
This suggests the antiferromagnetic nature of the phase transition.
\\
\begin{figure}[!t]
\centering
\includegraphics[scale=0.36]{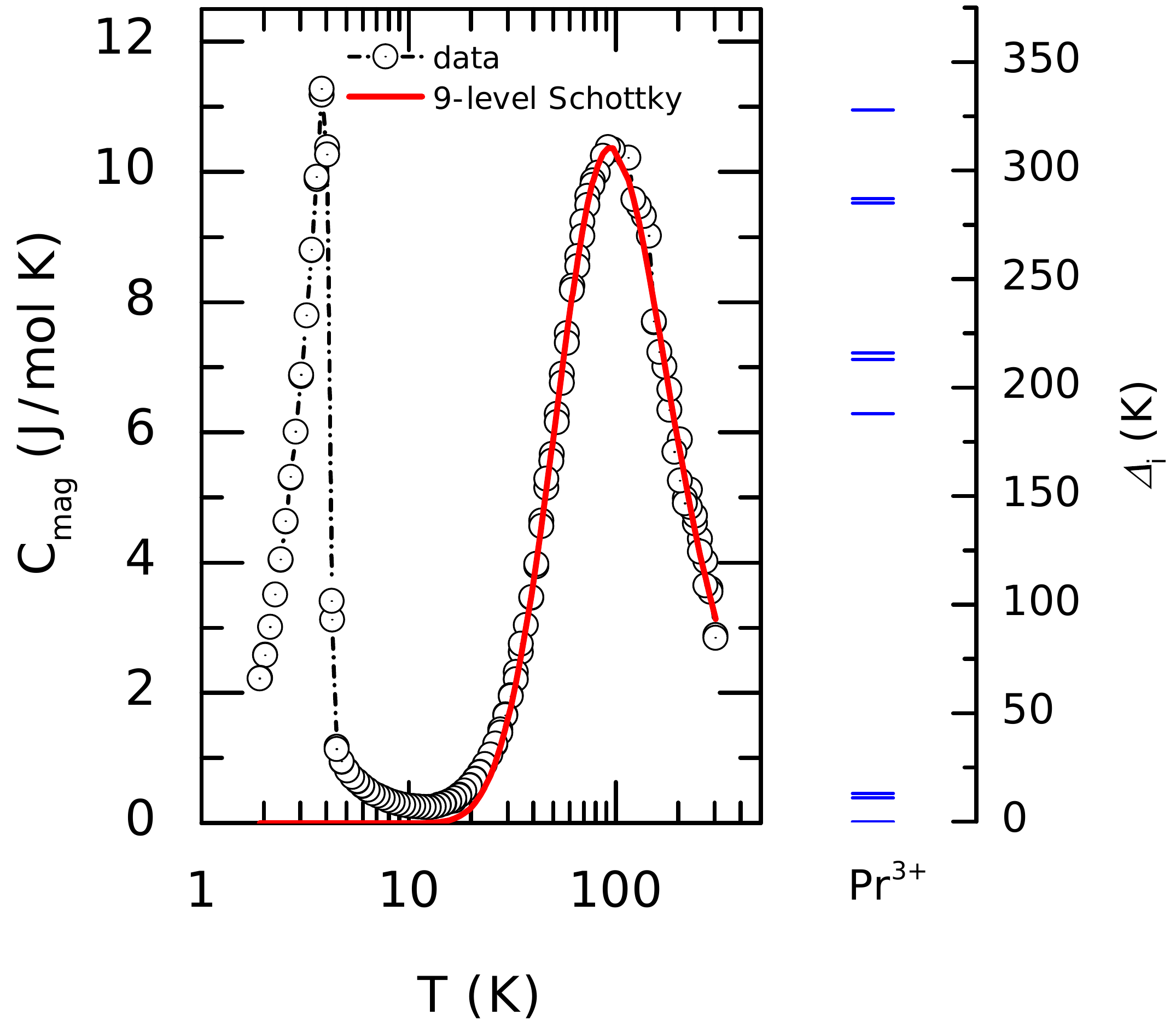}
\caption{\label{fig_schottky} The magnetic part $C_\mathrm{mag}$ of PrFe$_2$Al$_8$ is shown in black circles in a semi-log plot. The magnetic phase transition at $T_{N2}$ is clearly visible as a sharp peak. A Schottky-like peak is observed as centred at $\approx$70~K which is modelled by assuming a 9-level energy scheme [Eqn.~(\ref{eqn_schottky})]. The estimated CEF energy levels, $\Delta_i$, are shown.}
\end{figure}
\indent
Fig.~\ref{fig_chi} (a) shows the real part of ac susceptibility, $\chi'(T)$, 
measured at 10~Hz and at 100~Hz. Both the anomalies 
$T_{anom}$ and $T^{Pr}_{N}$ are reproduced. 
However, the ac susceptibility data do not present any 
frequency-dispersion for the peaks.
In order to confirm the nature of the peak at $T_{anom}$, 
$\chi'(T)$ was measured at more 
number of frequencies as shown in (b). However, very feeble 
frequency dependence of the peak was observed 
(from 32.99~K for 0.1~Hz to 33.50~K for 999~Hz)
which did not conform to response from spin glass-like 
clusters. Hence, if at all short-range static or 
dynamic magnetism arising from Fe-lattice is present in 
the sample of PrFe$_2$Al$_8$, it remains
masked by the long-range magnetism of Pr. Microscopic 
probes of magnetism like M\"{o}{\ss}bauer spectroscopy, NMR or
$\mu$SR are required to extract the Fe-magnetism in this case. 
As a final step in the analysis of the 
nature of magnetism at $T^{Pr}_{N}$, Arrott plots are 
presented in Fig.~\ref{fig_arrott} which is derived
from the magnetization isotherms in the range 2~K to 15~K. 
Notably, the
low temperature curves do not support a pure ferromagnetic 
feature where
the Arrott plots ought to be straight lines\cite{yeung1986arrott}.
Instead, an "S"-like curvature is present in (a) which could 
suggests first-order-
like nature or inhomogeneous magnetism. In the  case of (b), 
a close similarity with the 
case of a disordered ferromagnet is apparent\cite{yeung1986arrott}.
\subsection{\label{Cp} Specific heat}
\indent
The experimentally measured heat capacity, $C_p (T)$, of 
PrFe$_2$Al$_8$ is presented in 
Fig.~\ref{fig_cp} (a) plotted as circles. A clear peak in $C_p(T)$ 
signifying the phase transition 
at $T^{Pr}_{N}\approx$ 4~K is observed. As a non-magnetic reference, the $C_p(T)$ of 
LaFe$_2$Al$_8$ is presented in (a) as a solid line. The low temperature 
part of the specific heat 
is plotted as $C^\mathrm{diff}_{4f}/T$ versus $T^2$ where, 
$C^\mathrm{diff}_{4f}$ = ($C_p$(Pr) - $C_p$(La)) in the inset of (a) 
along with the linear fit (red solid line) which estimated the Sommerfeld coefficient, 
$\gamma_\mathrm{4f} \approx$ 6~mJ/mol-K$^2$. It is to be noted that
the value, $\gamma_\mathrm{4f}$ that we have obtained is through a subtraction procedure
and actually assumes comparable lattice specific heats for
the Pr and La compounds.
The overall feature of the $C_p/T$ versus $T^2$ in the present 
case is similar to that of EuRh$_2$In$_8$ where a transition 
at 5~K occurs first and further 
low in temperature, a broad anomaly occurs at 1.6~K
\cite{fritsch_prb_2004antiferromagnetic}. 
The magnetic specific heat obtained by subtracting the 
$C_p(T)$ of LaFe$_2$Al$_8$ from that of PrFe$_2$Al$_8$ 
displays a very broad peak centered at approximately 70~K 
(shown later in Fig.~\ref{fig_schottky}). The 
latter peak might arise from the crystal field levels of Pr$^{3+}$. 
The magnetic entropy, $S_m$, estimated 
through numerical integration of the magnetic specific heat is 
presented in Fig.~\ref{fig_cp} (b). Assuming the 
presence of Pr$^{3+}$ ($J$ = 4), the theoretical value of 
$S_m$ = $R$~ln(2$J$ + 1) is $\approx$ 18~J/mol K. At about 180~K 
this entropy is recovered while only about 40$\%$ of this value 
is attained at close to the magnetic transition at $T^{Pr}_{N}$. 
In Fig.~\ref{fig_cp} (b), $S_m$ attains a value of $R$~ln(2) at 
$T^{Pr}_{N}$ thereby confirming a doublet state for Pr$^{3+}$.
This observation together with the low and normal-metal 
value of 6~mJ/mol K$^2$ for the Sommerfeld 
coefficient show that the magnetic ordering at $T^{Pr}_{N}$ in 
PrFe$_2$Al$_8$ involves well-localized
magnetic moments of Pr$^{3+}$ ion. In the panel (c), a plot 
of $C_p/T^3$ versus $T$ is shown which brings 
out the phononic contribution in this compound. Such a plot 
would exhibit a sizeable peak at $T_{max}$ if significant contribution 
from optical phonons is present. In the present case, $T_{max}$ 
is quite broad and centred roughly around 30~K. For a comparison, 
the plot of LaFe$_2$Al$_8$ is also present which exhibited a 
similar feature while CeFe$_2$Al$_8$ and CeCo$_2$Al$_8$ 
did not\cite{ghosh_actapolonica_2012strongly}.
\\
\begin{table}[!t]
\centering
\caption{\label{tab3} The peculiarities of magnetic ordering in different Pr$T_2X_8$ compounds are collected. SRO = short-range order. LRO = long-range order.}
\setlength{\tabcolsep}{9pt}
\begin{tabular}{lll}\hline\hline
Pr$T_2X_8$            &     Pr-ordering            &     $T$-ordering \\ \hline
PrFe$_2$Al$_8$     &     4~K (LRO)      &      SRO may be present    \\
PrCo$_2$Al$_8$     &       5~K  (LRO)                   &      not observed down to 2~K             \\
PrFe$_2$Ga$_8$    &       15~K  (LRO)                 &      SRO may be present             \\
PrCo$_2$Ga$_8$    &     not observed down to 2~K                  &     not observed down to 2~K           \\ \hline\hline
\end{tabular}
\end{table}
\indent
The magnetic part of the specific heat obtained by subtracting 
the specific heat of LaFe$_2$Al$_8$ from that of PrFe$_2$Al$_8$ was 
used to analyze the crystalline electric field (CEF) levels of Pr$^{3+}$ ion. 
In the given symmetry of the space group $Pbam$, 
the Pr ion is situated at the Wyckoff position of $4g$ which has a 
point symmetry $m$. In this orthorhombic crystal field 
environment, Pr$^{3+}$ with $J$ = 4 would possess 9 singlet levels. 
The multilevel Schottky expression for a $^3H_4$ system 
with nine singlet levels is written as:
\begin{equation}
C^{CEF}_p = k_B\left[\sum_{i = 1}^{9}\left(\frac{\Delta_i}{k_BT}\right)^2p_i - \sum_{i = 1}^{9}\left(\frac{\Delta_ip_i}{k_BT}\right)^2\right]
\label{eqn_schottky}
\end{equation}
where $p_i$ = $Z^{-1}$ exp(-$\Delta_i$/$k_B$T) is the Boltzmann 
population factor, $Z$ = $\sum_{i}$ exp(-$\Delta_i$/$k_B$T) 
is the partition function, $k_B$ is the Boltzmann constant and 
$\Delta_i$ are the locations of the CEF energy levels above zero. 
A starting point for the crystal field energy-level values were 
obtained by modelling the case of Pr$^{3+}$ in 
PrFe$_2$Al$_8$ using a point-charge model employing the 
software McPhase and the subroutine so1ion\cite{mcphase}. The 
energy-level values obtained using this method were used in 
the above expression in order to perform a least-squares fit of the magnetic 
specific heat data shown in Fig.~\ref{fig_schottky}. Since the 
values of $\Delta_i$ from inelastic neutron 
scattering experiments are not available for PrFe$_2$Al$_8$ 
or closely related systems, tentative 
initial values were instead obtained from the reports on 1:1:3 
systems PrGaO$_3$\cite{senyshyn_jpcm_19_2007} 
and PrIrSi$_3$\cite{anand_arxiv_2014}. A curve fit to the 
experimental data following Eqn~(\ref{eqn_schottky}) 
is presented in Fig.~\ref{fig_schottky} and the estimated 
CEF energy levels, $\Delta_i$, are also shown. 
It must be noted that an accurate estimation of the CEF 
levels, $\Delta_i$, is best possible through 
inelastic neutron scattering methods and usually the Schottky 
contribution to the specific heat is calculated using the $\Delta_i$ 
thus derived\cite{anand_arxiv_2014}. The amorphous ferromagnetic 
contribution from the Fe sub-lattice would have very little 
contribution towards the total magnetic entropy however the 
exchange field generated this way may polarize the 
low-lying crystal field levels. The magnetic entropy plotted in 
Fig.~\ref{fig_cp} (c) clearly shows full doublet close to 
$T^{Pr}_{N}$ whereas the CEF analysis shows the first two 
excited energy levels at approximately 
10~K. An almost full occupation of 9 levels is attained at 
$\approx$ 200~K in Fig.~\ref{fig_cp} (c) whereas the CEF analysis 
presents highest level at above 300~K. These qualitative 
inconsistencies might have their origin in the subtraction 
procedure performed to obtain the magnetic specific heat, 
or inaccuracies in the estimation of 
crystal field levels. The distinct feature in $S_m(T)$ at 
$T^\mathrm{Pr}_{N}$ however is unequivocal and 
reproducible regardless of whether
$C_p(T)$ or $C_{4f}(T)$ is inspected for the entropy 
change at $T^\mathrm{Pr}_{N}$. Alternatively, the itinerant 
electron magnetism inherent to 
the material also can contribute towards such a feature. 
Keeping in mind that a faithful estimation of 
the CEF levels are possible through inelastic neutron 
scattering methods, our future endeavour would be to carry out an accurate 
estimation of the CEF levels.
\section{\label{DISCUSSION}Discussion}
The magnetic properties exhibited by Pr$T_2X_8$ compounds 
calls for deeper investigations involving microscopic 
structural probes like neutrons and $\mu$SR. The structural 
peculiarities comprising of the quasi cage-like
structure of Al polyhedra and the strand-like alignment of 
Pr and $T$ along the $c$-axis are of key significance. 
Among the various $RT_2X_8$ compounds investigated, the 
Ce-based compounds CeFe$_2$Al$_8$ and CeCo$_2$Al$_8$
displayed no signature of magnetic ordering down to 0.4~K
\cite{ghosh_actapolonica_2012strongly}.
These compounds were investigated for the strongly correlated 
electron behaviour. In CeCo$_2$Al$_8$, a stable 
Ce magnetic moment was observed while in CeFe$_2$Al$_8$, 
intermediate-valent states. Single crystal studies on
CeCo$_{2-x}M_{x}$Al$_8$ [$M$ = Mn, Fe, Ni] concluded that 
the $4f$ moment was not perturbed by the 
size of the dopant atom\cite{treadwell_ic_2014investigation}. 
However, signs of ferromagnetism emerged in 
the case of doping with Mn and Fe. A report on neutron 
diffraction investigation on CeFe$_2$Al$_8$ did not
reveal magnetic ordering down to 1.5~K\cite{kolenda_jac_2001low} 
however, the magnetization and M\"{o}{\ss}bauer 
data suggested clusters of Fe below 6.5~K. On the other hand, 
PrCo$_2$Al$_8$ was found to order
magnetically below 5~K\cite{tougait_jssc_2005prco} with a 
metamagnetic-like transition at a critical field of 0.9~T. 
Interestingly, the magnetic susceptibility was reported to 
adhere to ideal Curie-Weiss behaviour with an effective 
magnetic moment
of 3.48(5)~$\mu_\mathrm{B}$/f.u. suggesting the presence 
of stable Pr$^{3+}$
moment. No contribution to magnetism from Co moment 
was inferred. In Table~\ref{tab3},
the magnetic properties of the Pr-based $RT_2X_8$ 
compounds are collected.
\\
\indent
Rietveld analysis of the laboratory x ray data indicates 
that the Pr-based 1:2:8 compounds PrFe$_2$Al$_8$,
PrFe$_2$Ga$_8$ and PrCo$_2$Ga$_8$ belong to the same 
orthorhombic $Pbam$ space group  with similar lattice 
parameters and comparable interatomic distances for 
Pr--Pr, Fe--Fe distances (see Table~\ref{tab2}).
PrCo$_2$Ga$_8$ is found to display no signs of magnetic 
ordering (data not presented)
while PrFe$_2$Ga$_8$ shows a magnetic transition at 15~K. 
Given the structural similarities between these 
compounds, it is interesting to observe multiple magnetic 
anomalies in PrFe$_2$Al$_8$. Both the anomalies
do not display any frequency dispersion in ac susceptibility 
measurements while the anomaly 
at 4~K is suppressed upon the application of 10~kOe magnetic field. 
An added proof for the Pr ordering at $T^{Pr}_{N}$ is obtained
from the magnetic entropy derived from total specific heat 
(Fig.~\ref{fig_cp}) where $S_m$ = $R$ln(2) is recovered at $T^{Pr}_{N}$. However,
Arrott plots indicate the presence of inhomogeneous 
magnetism at low temperatures\cite{yeung1986arrott,das2014anisotropic}. 
Compared to the reported case of 
PrCo$_2$Al$_8$\cite{kolenda_jac_2001low}, the compounds 
of the present paper PrFe$_2$Al$_8$, PrFe$_2$Ga$_8$ and 
PrCo$_2$Ga$_8$ are seen to exhibit non-linear features in 
inverse magnetic susceptibility clearly suggesting the role 
played by the crystalline electric field effects. Most importantly, 
the 1/$\chi(T)$ data of PrFe$_2$Al$_8$ presented in Fig.~\ref{fig_cw} (a)
supports the role of CEF levels of Pr$^{3+}$ as well as 
the presence of magnetic fluctuations of Fe which persist
up to 300~K or above. 
\\
\indent
The specific heat and CEF analysis of Pr-based cubic systems 
are of fundamental interest to probe the resulting magnetic ground state.
Several prototypical compounds have been investigated in 
this regard. A number of Pr-based systems show magnetic order 
despite a nonmagnetic ground state suggested by the CEF predictions. 
In fact, magnetic order can be established
via the admixture of higher lying CEF levels into a ground state of 
higher degeneracy in the case of exchange energy larger than
the first CEF excitation\cite{bleaney1963crystal}. Such CEF assisted 
ferromagnetic ordering is observed in the case of 
PrNiGe$_2$\cite{snyman2013studying}. On the other hand, if the 
exchange interactions between the rare earth
ions dominate over the CEF, the magnetic moment associated 
with the $4f$ electrons can be completely 
quenched\cite{trammell1963magnetic}. Induced magnetism in the 
singlet ground state system PrIrSi$_3$
has been observed with a surprisingly large energy splitting of 92~K
\cite{anand2014investigations}. Another Pr-based 
compound, PrAu$_2$Si$_2$ and its doped variant 
PrAu$_2$(Si$_{1-x}$Ge$_x$)$_2$ are systems where 
dynamic fluctuations of the CEF levels were found to hold the 
key to destabilizing the induced moment magnetism
and realizing spin glass state\cite{krimmel1999spin,goremychkin2008spin}.
In PrPtBi, for example, Pr$^{3+}$ occupies a cubic local symmetry 
which splits the ninefold degenerate 
ground state of $^3H_4$ in to one singlet ($\Gamma_1$), one 
doublet ($\Gamma_3$) and two triplets 
($\Gamma_4$, $\Gamma_5$)\cite{suzuki1997non}. An energy 
separation of 87.5~K between the ground state and the first
excited state was estimated. 
\section{\label{CONCLUSION}Conclusions}
In conclusion, we have studied the magnetic and thermodynamic properties of the intermetallic compound PrFe$_2$Al$_8$ which crystallizes in $Pbam$ symmetry with quasi cage-like structure of Al atoms. The structural features of PrFe$_2$Al$_8$ resembles those of other isostructural compounds like LaFe$_2$Al$_8$ and PrCo$_2$Al$_8$. Magnetic measurements reveal two anomalies at $T_{anom} \approx$ 34~K and $T^{Pr}_{N}\approx$ 4~K. The latter anomaly is likely to originate from magnetic ordering in the Pr lattice however, a conclusion about this requires further experimental proofs. Signatures of short-range magnetic fluctuations in the Fe lattice are also obtained however, could be mediated by impurities. Metamagnetic-like steps in the magnetization isotherm at 2~K is clearly evidenced. The phase transitions at $T^{Pr}_{N}\approx$ 4~K is also reflected in the specific heat. Analysis of magnetic entropy identifies a value of $R$~ln(2) at 4~K suggesting the presence of a Pr doublet crystal field level in spite of the symmetry of Pr site predicting only 9 singlet levels. Neutron powder diffraction and inelastic neutron scattering experiments are called for to understand the magnetic properties and the CEF level schemes of PrFe$_2$Al$_8$.
\\ \\
$\ddagger$Present address: Department of Applied Physics, Birla Institute of Technology, Mesra, Ranchi, Jharkhand, India
\\
\section*{Acknowledgements}
H. S. N. and R. K. K. acknowledge Manh Duc Le for help with the McPhase software and CEF calculations. 
They also thank the FRC/URC for a postdoctoral fellowship. A. M. S. thanks the SA-NRF (93549) and the FRC/URC of UJ for financial assistance. 
\\
\\
%
%
%\bibliography{RT2X8}
%\bibliographystyle{apsrev}
%

%
\end{document}